# Ageing in Which Place? Spatial Analytical Framework for Evaluating Ageing-in-Place Practices

Yong Tu[1], Yao-pei Wang[2], Yumeng Yang[3], Yi Fan[4]

[1] Department of Real Estate, School of Business, National University of Singapore, 15 Kent Ridge Drive, Singapore 119245, Singapore. Email: biztuy@nus.edu.sg. RCID ID: 0000-0002-4581-3676.

[2] Chinese Academy of Housing and Real Estate, Zhejiang University of Technology, Hangzhou, China. Corresponding author and e-mail: wangyaopei123@163.com. ORCID ID: 0000-0002-6729-7780.

[3]The State Information Centre, Department of Big Data Development, 100045. yangyumeng@sic.gov.cn.

[4] Department of Real Estate, School of Business, National University of Singapore, 15 Kent Ridge Drive, Singapore 119245, Singapore. Corresponding author and email: yi.fan@nus.edu.sg. ORCID ID: 0000-0001-6530-8282.

**Author contributions**

**Yong Tu:** Conceptualization; Supervision; Resources; Writing – Original Draft; Funding Acquisition

**Yao-pei Wang:** Writing – Original Draft; Data Collection and Analysis; Methodology; Visualization

**Yumeng Yang:** Data Collection and Analysis; Methodology

**Yi Fan:** Writing – Review & Editing; Visualization; Project Administration; Supervision

**Declaration of interest:** The authors have no competing interests to declare that are relevant to the content of this article.

**Fund:** We appreciate the financial support from the National University of Singapore Academic Research Fund (A-8002711-00-00).

**Data availability**

The data were extracted from OneMap (www.onemap.gov.sg), an online map of Singapore that provides the most detailed and timely spatial information, and an official online database (www.data.gov.sg) that provides socioeconomic and population data. Both are freely available to the public.

# Ageing in Which Place? Spatial Analytical Framework for Evaluating Ageing-in-Place Practices

## Abstract

Over the past decade, governments around the world have made significant investments in creating elderly-friendly urban environments within local neighborhoods. However, the lack of a standardized evaluation framework for Ageing-in-Place (AIP) practices makes it challenging to generalize these experiences. First, we compare the AIP models of the U.S.-San Francisco, Japan-Tokyo, and Singapore using a cost-benefit analysis, demonstrating the comparative advantage of the Singapore model in terms of low cost and high accessibility for the independent ageing population. Second, we propose a spatial analytics framework to visualize and quantify the degree of alignment between a basket of ageing facilities and the active ageing population, enabling a data-driven, timely evaluation of the effectiveness of Singapore's AIP policies. Singapore's AIP model, either in its entirety or as a hybrid with other models, can be generalized to other global cities, providing valuable insights for optimal elderly-friendly urban planning.

## Introduction

Increasing longevity is a significant achievement of modern societies, but it burdens shrinking workforces and strains social care systems (Yang, 2019). By 2030, 1 in 6 people worldwide will be aged 60 years or over. This number is expected to double by 20250, according to the World Health Organization (WHO). A longer life brings more opportunities, not only for older people but also for their families and communities. However, the proportion of people living in good health has remained broadly constant, implying that additional years are often associated with poorer health (Scott et al., 2021). What can urban planners do to help older people stay active and increase their healthy lifespan? How can we manage the costs of an ageing population while ensuring that older people maintain a decent quality of life? We endeavor to answer these questions in the context of ageing-in-place (AIP) practices and propose a multistage urban spatial analytic framework to address this issue.

Due to the diverse geographical, economic, and cultural contexts across countries (Huang et al., 2025 & 2022; Yang, 2019; Makizako et al., 2021), AIP practices vary significantly across different contexts, though few studies have conducted comparative analytics of different AIP models using standardized criteria (WHO, 2015; Shinan-Altman, Gum & Ayalon, 2020). By comparing the AIP practices in the U.S.-San Francisco, Japan-Tokyo, and Singapore under a cost-effective framework, we find that the Singapore model is conceptually more inclusive to the different types of older people—*dependent and independent (active ageing)* elderly—than the U.S.-San Francisco and Japan-Tokyo models, and financially more feasible for the older people. This is attributed to the Singapore government's carefully planned and top-down AIP policies, which invest considerable centralized funds in boosting neighbourhood ageing facilities and cares.

Most existing studies on AIP remain conceptual or theoretical (Pani-Harreman et al., 2021; Golant, 2020), with limited empirical evaluations of AIP practices (Moradi, Biloria & Prasad, 2023; Sulis & Proietti, 2024), making it difficult to assess the efficiency of an AIP model. Specifically, prior theoretical studies lay out conceptual frameworks, such as the 'Development Approach,' 'Social-technological Innovation Approach,' and 'Not a One-Size-Fits-All Concept' (see details in the next section), particularly the Public and Private Partnership in Delivering Affordable and Sustainable AIP Approach which was proposed by the WHO (2012 & 2015) and emphasizes the long run affordability and sustainability of an AIP practice. However, there are limited empirical evaluation and validation on these models. Most empirical literature in this area focuses either on the dependent or disabled elderly (Kim & Koh, 2022; Wang, Yang & Avendano, 2022; Qiu, Zhang & Cheng, 2023), or on a specific type of ageing-friendly facility, such as the impacts of green space or health care facilities on ageing population (Chen & Chang, 2015; Rekha et al., 2017). Fewer studies investigate how an AIP model improves the quality of life of *active ageing* older adults' practices (Klicnik et al., 2024).

Being different from the existing literature, this paper aims to adopt empirical methods to holistically evaluate an AIP model in an urban area. We firstly conduct a comparative cost–benefit analysis of AIP practices for three representative models drawn from the AIP practices in the U.S., Japan, and Singapore. Both the U.S-San Francisco and Japan-Tokyo AIP practices focus more on providing services and facilities to *dependent* older people, while overlooking the needs of *active ageing* seniors, while the Singapore AIP policies are centred on creating an ageing friendly built environment, such as a neighbourhood, in which both active and vulnerable seniors can live together with people from other age groups. Nevertheless, globally more than 80% of the ageing population are independent and can exercise self-care. This analysis demonstrates how different AIP frameworks can deliver different outcomes in ageing care, advocating the importance of evaluating and conceptualising AIP practices.

Secondly, we propose an adaptable framework that quantifies the spatial matching between elderly-friendly facilities and the distribution of the older population in Singapore. This framework consists of two essential steps. First, we employ a spatial network analysis that uses a gravity model to estimate the accessibility of facilities for older people within a specified time distance. This analysis allows us to visualize the spatial patterns between facility distribution and the older population. Next, we develop an *AIP Facility Matching Index* (AFMI) that provides a quantitative measure of the degree of spatial matching. This index illustrates whether there is an oversupply or shortage of AIP-friendly facilities. The findings from our network analysis and AFMI scores reveal instances of facility oversupply or shortage in specific planning areas. Methodologically, this framework can visualize the mismatch between ageing population and ageing friendly facilities at a broad city/region level, guiding policy efforts in optimizing resource allocation from the top to ensure equity across regions. We therefore contribute to the quantitative evaluation of the AIP model. Our spatial model can help reconceptualize the current AIP framework, by taking the spatial (mis)match into account.

Specifically, by introducing Singapore's AIP practice and proposing an adaptable framework to enhance implementation efficiency, our paper makes the following contributions. To the best of our knowledge, we are the first to apply the gravity model derived from the spatial network analysis to the context of active ageing and estimate accessibility to a comprehensive range of elderly-friendly facilities, explicitly quantifying the spatial balance between AIP facilities and the older population (Huang et al., 2024 & 2021; Sulis & Proietti, 2024). Literature on the spatial (mis)match between ageing facilities and the population of the older people is thin, with most work focusing on the dependent or disabled elderly (Kim & Koh, 2022; Wang, Yang & Avendano, 2022; Qiu, Zhang & Cheng, 2023), or a specific kind of facility, such as green space (Chen & Chang, 2015) or healthcare resource (Rekha et al., 2017). In terms of the AIP practice, the Singapore's AIP model introduces an effective way to support older people's ageing needs by providing public goods of affordable facilities and services within communities, allowing them to stay in the neighborhoods as long as possible.

Methodologically, we are able to derive a comprehensive set of data on place attributes from publicly available websites of ten government authorities in Singapore.[1] Additionally, Singapore's public housing communities, comprising of nearly 80% of the population, are highly homogeneous and ungated, which helps us eliminate any unobserved confounding factors, if any, arising from otherwise variation in the community facilities. Thus, we are able to focus on the variation from the accessibility of AIP facilities across communities, as an important measure of disparities in AIP services. In sum, this study provides an evaluation framework to measure the spatial distribution balance between older population and the AIP facilities. It can guide policy efforts in their age-well planning and shed light on optimal elderly-friendly urban planning to support sustainable ageing.

## Ageing in Which Place? A Comparative Analysis of AIP Practices

### *Comprehensive understanding of AIP*

AIP originated from environmental gerontology. Previous research has identified four key approaches along the evolution of the AIP concept to guide the successful implementation of AIP in practice. The initial "Development Approach" emphasizes improving seniors' mental wellbeing (Erikson & Erikson, 1998). Subsequently, the "Public and Private Partnership in Delivering Affordable and Sustainable AIP Approach" offers a framework for sustainable AIP, ensuring the affordability of services and facilities for older people in the long term without placing an unsustainable burden on public spending (WHO, 2012, 2015). The "Social-technological Innovation Approach" (Heinze & Naegele, 2012) recognizes the importance of integrating social or policy innovation with technology. Lastly, the "Not a One-Size-Fits-All Concept" underscores the need for context-sensitive AIP practices that account for individual differences, cultural variations, temporal factors, location specifics, and socioeconomic context (Bigonnesse & Chaudhury, 2022).

[1] The government authorities include Ministry of Health (MOH), Ministry of Education (MOE), Ministry of Manpower (MOM), Housing Development Board (HDB), Ministry of Family and Social Development (MSFD), Ministry of Transport (MOT), Ministry of Culture Community and Youth (MCCY, including people's Association), Urban Redevelopment Authority (URA), Ministry of National Development (MND), and National Parks (NParks).

We adopt the definition of AIP proposed by Golant (2015, 2020) in the literature of environmental gerontology that "*Older people grow old in some situational, contextual, or environmental vacuum. It is as if the dwellings, buildings, neighborhoods, communities, and regions in which they live and their built, natural, social, organizational, and political environments make little difference in whether they enjoy their lives, feel good about themselves, live independently, and achieve healthy lifestyles*". It serves to alleviate the burden on institutionalized care and aligns with the preference of self-care seniors (Forsyth & Molinsky, 2021). Additionally, the socioemotional selectivity theory explains that as people age, their motivations related to social belongingness become stronger, and they experience higher levels of satisfaction when living in familiar homes and neighborhoods (Geboy et al., 2012; Golant, 2020). Therefore, facilitating independent older people, who account for over 80% of the total elderly population in urban areas, to age in their own neighborhoods is an ideal strategy for AIP practice (Rowe & Kahn, 2015; Repo, 2024). Ageing-in-place aims to connect older people with environments that offer high levels of physical, psychological, and social functionality, enabling the older people to maintain a certain degree of independence (Bigonnesse & Chaudhury, 2022). The creation of an age-friendly built environment holds significance not only for older people's quality of life, but also for their families and institutions responsible for providing services and care (Wahl et al., 2012; Sulis & Proietti, 2024).

While literature generally agrees that "place" is defined as older people's own homes or local communities (Yu & Rosenberg, 2017; Pani-Harreman et al., 2021), it is more than a physical location. The place encompasses physical, social, and psychological dimensions that can impact on older people's behavior, social interaction, and lifestyle. In the physical dimension, the urban physical environment not only increases the functional capacity of older people but also extends their independent period (Wang et al., 2021). From the social perspective, literature shows that social isolation accelerates the ageing process (Bzdok & Dunbar, 2022; Qi et al., 2023). The AIP practice encourages active ageing older people to interact at public space, resulting in a higher quality of life (Emmer De Albuquerque Green, 2022), greater benefits and socioeconomic advantages than simply prolonging lifespan (Scott et al., 2021). It also aids dependent older people to age at their own home and neighborhood. From the psychological dimension, more outdoor activities are found to improve people's mental health along the ageing process (van Hees et al., 2017; Zhang et al., 2023).

In the AIP context, social institutions are found to have significant effects on the well-being of older people (Goldman et al., 2018), calling for collective actions and government interventions. Existing AIP practices pay more attention to narrowly-defined groups of elderly, such as dependent or half-dependent seniors, leaving the fully independent seniors less cared for. Guided by the theoretical and conceptual discussions of the AIP practice, the urban planners should improve the AIP facilities in the local vicinity for independent elderly too to create an age-friendly environment to support their healthy ageing in place.

*The U.S-San Francisco Model: Continuing care retirement communities*

Table 1 shows the primary way the United States provides AIP facilities—by establishing a specific type of AIP neighbourhood called continuing care retirement communities (CCRCs). While retirement options in the United States are diverse, including supportive service programs in naturally occurring retirement communities (NORC-SSPs), village projects (such as Beacon Hill), simple congregate living arrangements (both affordable and market rate), active living communities, senior cohousing, and freestanding assisted living, CCRCs represent the highest proportion among all types of AIP neighbourhoods (Shinan-Altman et al., 2020). In fact, nearly 55% of the U.S. population resides in home-based retirement, but the government does not generally construct AIP facilities within regular communities.

Unlike standalone assisted living or nursing home facilities, CCRCs offer a comprehensive range of housing, health care, and support services in one location.

While some charitable organizations and faith-based groups may provide funding for the construction and operation of CCRCs, the government does not allocate any financial resources to them. The government does, however, exercise regulatory supervision and scrutiny of CCRCs and grants tax exemptions for CCRC-related expenses. Consequently, older people are responsible for bearing the living costs of CCRCs, which may include a lump-sum entry fee (charged by some CCRCs) and relocation expenses. Notably, the U.S. government has established Medicare insurance to assist older people with health care expenses (Kim & Koh, 2022). However, Medicare insurance does not cover the costs associated with CCRCs or home care services.

*The Japan-Tokyo Model: Government-funded home care and support services*

The Japanese government's current AIP policy is the implementation of long-term care insurance, which provides subsidies for home care expenses to help older Japanese people age at home. However, more than 90% of older people are active and independent in Japan and rely little on home care services. In fact, these older people are often active in outdoor activities and require AIP facilities in communities. Notably, the Japanese government implemented the "*Gold Plan*" in 1989 to build AIP facilities and develop community-based services for older people. In 2000, the *Gold Plan* was replaced by the long-term care insurance system. The Japanese government had two reasons for abandoning the *Gold Plan*. One lay in a lack of optimal planning and excessive cost. The other was manpower shortages and coordination issues in various functional departments during the large-scale urban regeneration process, which we will revisit in Singapore's context.

*The Singapore Model: Ageing for all older people at local communities*

The Singapore government relies on AIP planning, which is accessible and affordable to older people, with a focus on building AIP facilities within each community to support active ageing. In 2015, it announced the implementation of the '*Action Plan for Successful Ageing*', investing 3 billion to build more than 450 AIP facilities in 5 years.

In 2020, the '*Action Plan for Successful Ageing 2.0*' was announced, with the Singaporean government committing to further increasing investment in AIP facilities.

*Comparison and Conclusion*

A cost–benefit analysis was conducted to compare AIP practices in the U.S-San Francisco, Japan-Tokyo, and Singapore (see Table 1). It revealed that the CCRC model in the U.S-San Francisco imposes minimal financial burden on the government but does not necessarily improve welfare among older citizens and may increase inequality due to disparities in household wealth (Abeliansky et al., 2021). Moreover, in a survey conducted by the American Association of Retired Persons in 2021, over 77% of older people in the U.S. expressed a desire to age in their own homes. However, the reality is that conventional communities lack adequate AIP facilities. Japan's AIP model covers a larger proportion of older people under AIP welfare, but it focuses primarily on providing medical services to dependent or partially-dependent older people and overlooks the needs of independent older people. From the residents' perspective, Singapore's AIP model provides equitable access to high-quality AIP facilities for every active older person. Additionally, it offers transfer payments to households in the dependent group. Considering both government and resident perspectives, Singapore's practices appear to align best with the original purpose of AIP.

Table 1. Comparative cost–benefit analysis of AIP practices

| Cases | **The U.S-San Francisco Model** | **The Japan-Tokyo Model** | **The Singapore Model** |
|---|---|---|---|
| Groups receiving financial support | Dependent older people + Low-income older people who live in public housing communities | Dependent older people + Low-income older people who live in public housing communities | **Independent older people** + dependent older people + Low-income older people |
| Costs for government | **Approximately 9.78 dollars per senior per year**<br>Note: CCRC-related consumption is tax-exempt. | **Average 8.8 U.S. dollars per senior per year**<br>Note: Subsidies for home care expenses to assist Japanese population in AIP. However, approximately 90% of independent older people rely little on home care services. | **Average 196.6 U.S. dollars per senior per year**<br>Note: Government aims to invest 3 billion to build AIP facilities in each neighbourhood during 5-year Action Plan for Successful Ageing. |

| Costs for families | **Direct costs:**<br>**1. Lump-sum entry fee:** from $40,000 to more than $2 million;<br>**2. Monthly charge:** average $3,555 in 2021;<br>**Indirect costs:**<br>Relocation cost is approximately $20,000. | **Direct costs:**<br>**Home care costs:** average 26.4 U.S. dollars per senior per month;<br>**Indirect costs:**<br>Approximately 10 dollars per senior per month to commute to AIP facilities | **Direct costs:**<br>**Home care costs:** average 14.8 U.S. dollars per senior per month<br>Note: Singapore's 'ElderShield' insurance for seniors covers 30%–80% of home care service costs. |
|---|---|---|---|
| Benefits for older people or families | Long-term care services and facilities in retirement communities | Reduced payments for home care services | Convenience for active independent ageing population;<br>Reduced payments for home care services |
| Cost–benefit analysis for families | **High costs** with **limited accessibility** of AIP facilities and services | **Low costs but few AIP facilities** and services accessible to independent seniors | **No cost** for independent seniors and **low costs** for dependent older people, **high accessibility** of AIP facilities and services |
| Cost–benefit analysis for government | **Low costs**, 5%–10% of the older population enjoy the benefits of AIP practices | **Low costs**, approximately 20% of the older population enjoy the benefits of AIP practices | **High costs**, almost all older individuals enjoy the benefits of AIP practices |
| References and data sources | 1. United States of America: health system review (Rice et al., 2013).<br>2. https://www.aarp.org/c-aregiving/basics/info-2017/continuing-care-retirement-communities.html<br>3. https://www.businessn-ewsdaily.com/15842-costs-of-employee-relocation.html | 1. Estimated based on the data in Makizako et al. (2021).<br>2. Estimated based on the data in Zheng et al. (2022). | 1. The Singapore senior population is 637 thousand. We assume that the depreciation period for an AIP facility is 30 years.<br>2. https://www.moh.gov.sg/d-ocs/librariesprovider6/resources/eldershield-review-committee-report.pdf |

Figure 1 illustrates the features, costs, and benefits of the three AIP patterns across the U.S-San Francisco, Japan-Tokyo, and Singapore models. Different urban spatial structures, such as compact and noncompact types (Figure A2 in Supplementary Section 4), are suitable for different AIP patterns. As compact cities lack extensive space to build CCRCs and supplemental nursing homes, the Singaporean model provides a viable template for these cities. For noncompact cities, the government can also adopt a hybrid model, if the budget permits, to expand the coverage of elder care benefits. The analysis shows that different AIP practices produce different ageing-care outcomes, for example, the US-San Francisco and the Japan-Tokyo models incur low lost to the government but high cost to families with 5% to 20% of the older population benefiting from the AIP practices, while the Singapore model shows otherwise with almost all older population benefiting from its AIP practice. The findings motivate us to propose a spatial analytical framework to evaluate the efficiency of an AIP practice aiding future urban planning.

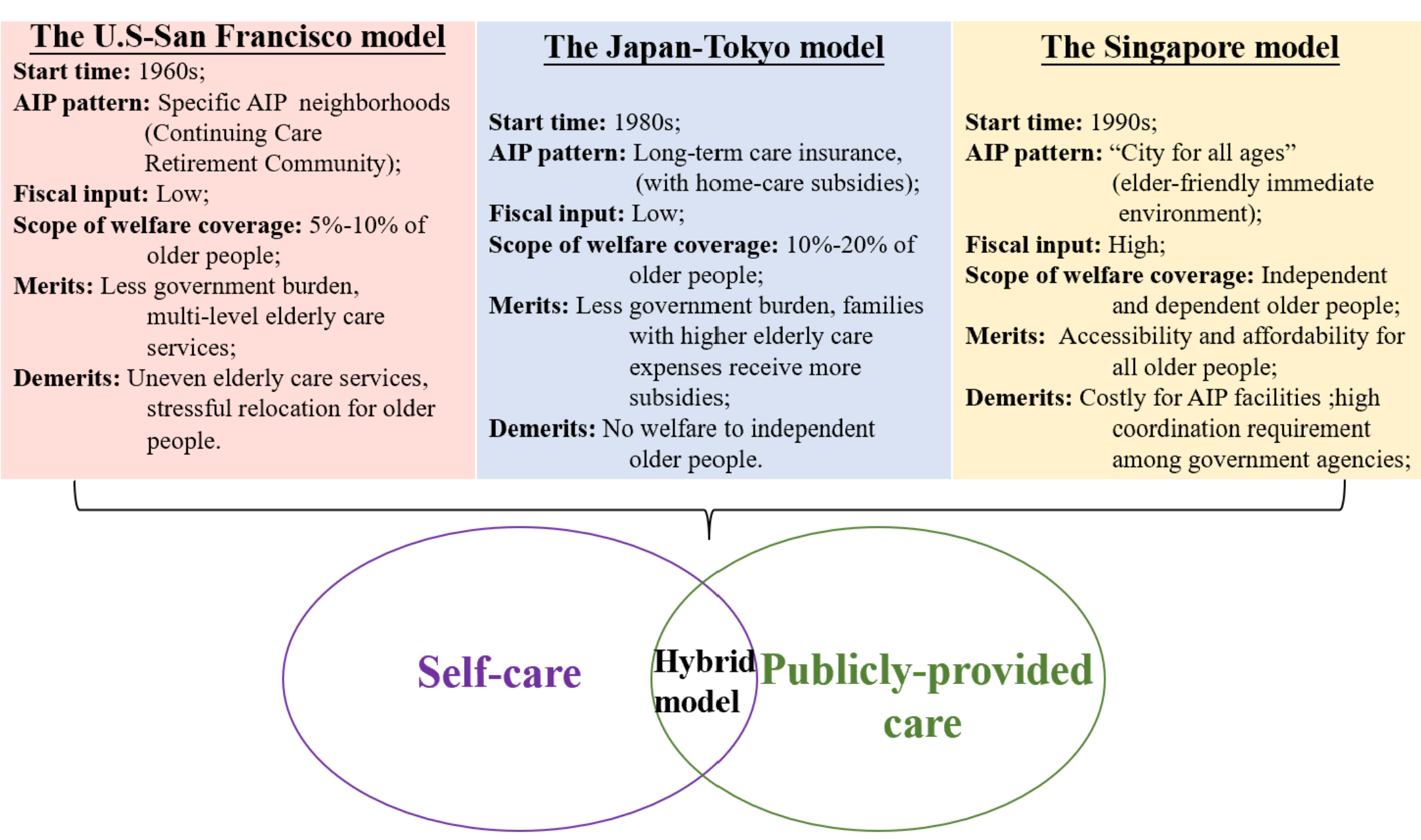

Figure 1. Comparative case analysis of AIP practices

**From Comparative Models to Spatial Analysis：logical connection**

The three selected case-study cities—San Francisco, Tokyo, and Singapore—represent distinct types of globally relevant AIP models. San Francisco illustrates a market-driven model that emphasizes self-funded elderly care services. Tokyo demonstrates a government-led model with a strong focus on home-based care and healthcare support.

Singapore exemplifies a state-coordinated, spatially efficient model that integrates planning across neighborhoods. Together, these cases provide a meaningful comparative basis for evaluating spatial strategies across diverse institutional and urban contexts. Following our detailed cost-benefit analysis from both governmental and resident perspectives, we gain insights into a broad spectrum of AIP governance and financing structures.

The Singapore model exhibits strengths in both the provision of physical facilities and the broader institutional dimensions of ageing support. Moreover, the model also excels in cultivating a psychologically and spiritually supportive environment. Ageing in place—within familiar surroundings and established social networks—strengthens older adults' sense of belonging and fosters sustained social engagement. In addition, co-residence with family members provides greater access to emotional intimacy and spiritual care, thereby reinforcing the holistic significance of 'place' within the ageing-in-place discourse.

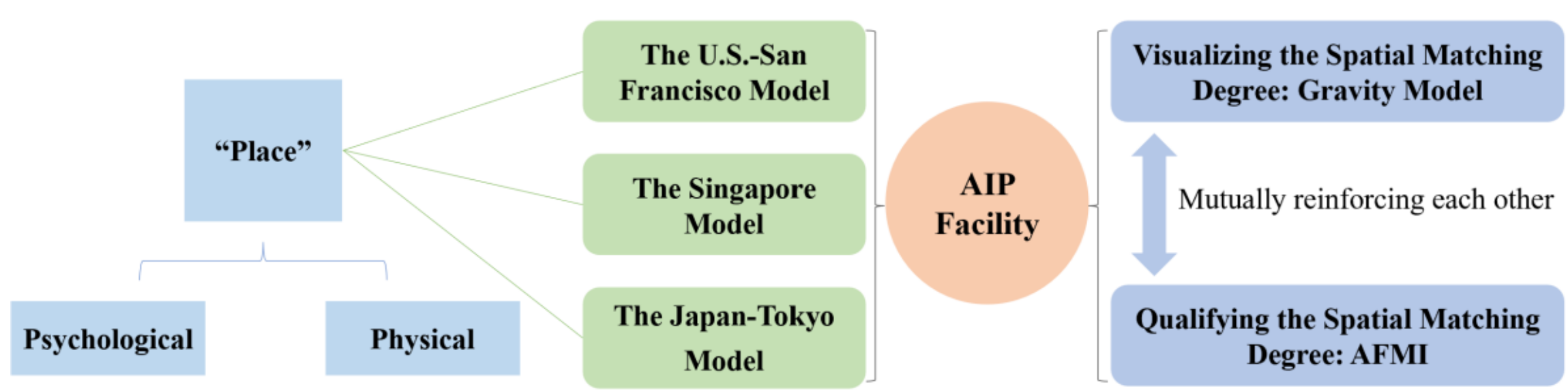

Figure 2. The connections among the theoretical background, the comparative analysis, and spatial analytic framework

Therefore, we recommend the promotion of the Singapore AIP model through the strategic development and distribution of elderly-care facilities, aligned with each country's existing institutional framework. However, as emphasized in our cost-benefit analysis and exemplified by Japan's experience, the effective implementation of this model demands government investment and long-term commitment. Ideally, AIP facilities should be equitably accessible to residents of all ageing cohorts within their respective communities. Nevertheless, variations in facility requirements across ageing cohorts can lead to spatial mismatches if the development process is not carefully coordinated. Despite these mismatches, the Singapore model demonstrates superior integration and internal consistency, making it a promising option for broader implementation, whether on its own or integrated into hybrid frameworks. While such hybrid models may be built on common assumptions, the persistence of spatial mismatches continues to pose a challenge to theories of spatial equity.

To effectively translate this model into practice, it is crucial to equip urban planners with robust frameworks for assessing spatial efficiency. Developing such tools is essential for supporting evidence-based decision-making, and this constitutes the primary focus of the next section of our study.

Figure 2 presents a schematic diagram that bridges the logical connections among the theoretical background, the comparative analysis, and the spatial analytical framework. The spatial analytical framework employs two complementary "absolute vs. relative" tools: the Gravity Model provides an absolute measure of accessibility, while the AFMI offers a relative measure of mismatch against a benchmark. Their distinction and synergy jointly capture the spatial alignment between AIP facilities and the elderly population. Specifically, the Gravity Model offers *absolute, facility-level measures* of accessibility—for example, the average number of elderly-friendly services accessible within 10 minutes across planning areas, as shown in Figure 3. In contrast, the AFMI provides *relative rankings* of planning areas based on the degree of mismatch between ageing facilities and the elderly population, benchmarked against the national average (Figure 5). This two-pronged approach is also consistent with prior studies (e.g., Ning et al., 2022), offering complementary perspectives from both absolute and relative dimensions and enabling a more comprehensive assessment of facility distribution and accessibility.

## Spatial Analytical Framework for Evaluating Ageing-in-Place Practices

The Singapore model, being the closest to achieving the goals of accommodating both active ageing population and dependent older people among current AIP practices, also has the largest government investment. Therefore, to achieve the equality and efficiency of an AIP practice and to promote the Singapore model, an urban spatial analytical framework is proposed to help governments optimize the equality and efficiency of AIP facilities and maximize the welfare of their older populations (Moradi et al., 2023). We thus use the Singapore model as a case study for this evaluation and decision-making process.

### *Data Collection*

The list of physical facilities needed by older people is based on WHO guidelines (WHO, 2007) and is subject to the latest availability. The location of and information on each facility were collected from different government websites, categorized into three groups and geocoded, as presented in Supplementary Information Table A1.

Geographic information on traffic networks in 2020 was matched to residential buildings, streets, bus routes, MRT lines, and footpath networks. These were used to calculate the time distance, which is the shortest time required to travel between a residential building and a specific ageing facility. In principle, a path reflects intermodal transportation and accounts for riding, walking, and waiting if scheduled public transport is used between the origin and destination (Hadas, 2013). However, this study does not consider waiting time because Singapore residents can easily track the arrival time of public transport from home. Finally, public residential buildings are scattered across 31 of Singapore's 55 planning areas, which serves as a blueprint for urban planners in designing urban activities. Population data on older people, defined as those aged 65 years and above, were extracted from the official online database in 2020.

### *Methods*

The urban spatial analytic framework employed in this study consists of two parts. The first one involves a spatial network analysis using the gravity model to visualize the spatial matching distribution between AIP facilities and the elderly population. The second part develops the AIP facility matching index (SMI) to quantify the degree of spatial matching.

To begin, we apply the network analysis method to examine the degree of spatial matching in Singapore's AIP practice within a specific planning area. The gravity model (Liu & Kwan, 2020) is employed to analyze the transportation network data and the distribution of facilities. This model allows us to estimate the number of facilities that a senior can access within a predefined time-distance. The gravity model is defined in Equation (1).

$$Gravity_i^r = \sum_{d[i,j]\leq r} \frac{W_j}{e^{\beta d[i,j]}} \quad (1)$$

where $i$ indicates older people's location '$i$' and $j$ indicates facility '$j$'. The $r$ is a time radius around residential location '$i$', which is assumed to be 10 minutes' time-distance for a single trip without considering the time needed to go down/up the staircase. $d[i,j]$ is the shortest time-distance between residential building '$i$' and facility '$j$', calculated using geocodes, and $W_j$ is the weight assigned to facility $j$. In this study, we assume $W_j$=1 for all $j$. The $\beta$ is the exponent that controls the effect of distance decay on each shortest path between $i$ and $j$. More specifically, $\beta$ is the weight derived from the speed of travel. The faster the travel speed, the larger the weight assigned to facilities locating farther away from the residential building (though within the time radius). It is worth mentioning that in principle the specification of $\beta$ depends on the mode of travel assumed in the analysis (e.g. walking, cycling, private driving, public transport). However, in this study, what we calculate is the nearest time-distance to AIP facilities under any mixed mode of travel. The default value of $\beta$ with hybrid travel is 0.00217 in ArcGIS.[2] As older people are slower in mobility, we chose a slightly larger $\beta$ (0.003) than the assumed value in ArcGIS, conditional on the standard walking speed of a typical adult and the limited mobility of the elderly. Literature shows that the average walking speed of adults in general is 1.1 to 1.4 m/s (Huijben et al., 2018) and the average walking speed of elderly people is lower than that of the general population (Duim et al., 2017). With no consensus on the average walking speed of elderly people across different countries, we assume that their walking speed is about 70% of that of adults in general, which is approximately 0.8 m/s (0.8/1.4 = 0.57 ~ 0.8/1.1 = 0.72). Therefore, the default value of $\beta$ was adjusted from 0.00217 to 0.003 (0.000217/0.7≈0.003). $Gravity_i$ is a nonnegative number. A larger value of gravity means that more age-friendly facilities are available within the assumed time radius of $r$.

[2] The equivalent value of β for distance units in "meters" is 0.00217; in "feet" 0.000663; in "kilometers" 2.175, and in "miles" 3.501.

In addition, the degree of spatial matching is quantified by an AIP facility matching index. The spatial mismatch theory originates from the analysis of the labor market (Kain, 1968), and now there is a branch that studies the mismatch of different public facility spaces, e.g., public schools (Sun et al., 2021) and urban parks (Chang & Liao, 2011). However, considering the spatial imbalance of AIP facilities, to the best of our knowledge, we are the first to adopt and develop the spatial mismatch theory to this field. The equation of AFMI is shown as follow:

$$\mathrm{AFMI}_{L,J} = \left| \frac{\left(\frac{h_{L,J}}{H_J}\right) P(Seniors) - P_L(Seniors)}{2P(Seniors)} \right| \tag{2}$$

where $AFMI_{L,J}$ stands for the spatial matching index in the $L^{th}$ planning area, *'L', L=1,…, n, n=31*, for the $J^{th}$ type of facility, with *J=0* indicating all facilities (see Table 1 for types of facilities). $P(Seniors)$ is the number of seniors at the national level, and $P_L(Seniors)$ is the number in planning area $L$. $H_J$ is the number of *J-type* facilities at the national level, and $h_{L,J}$ is the number of *J-type* facilities in planning area *L*. It is worth noting that Singapore is a highly homogeneous city-state, where external environmental characteristics are relatively consistent across its territory. As a result, the spatial distribution of older adults and elderly facilities is less affected by intra-city variation, making the use of national-level data as the baseline for the index both appropriate and reliable.

The absolute value of AFMI ranges between 0 and 1/2. A higher AFMI indicates a higher degree of spatial mismatch between seniors and facilities. A national average AFMI equal to 0 or 1 indicates a perfect match or mismatch, respectively (Kain, 1968; Holzer, 1991; Martin, 2004; Gobillon et al., 2015). The actual value of an AFMI, $\frac{\left(\frac{h_{L,J}}{H_J}\right)P(Seniors)-P_L(Seniors)}{2P(Seniors)}$, ranges between -1/2 and +1/2, with a negative value indicating facility shortage, zero indicating equilibrium, and a positive value indicating facility oversupply.

## Results

### *Visualizing Spatial Matching: Gravity Model*

Ensuring equitable distribution of AIP facilities across neighbourhoods requires significant financial investment. A crucial step in evaluating the effectiveness of AIP practices is assessing whether the spatial distribution of these facilities meets the demands of older people (Davern et al., 2017; Teow et al., 2018). We start with a network analysis to examine whether the distribution of AIP facilities satisfied the needs of the older population.

According to the WHO, a society is called an "*aging society*", "*aged society*", or *"superaged society*" when the proportion of people aged 65 and above exceeds 7%, 14%, or 20%, respectively. The current global trend of ageing is dramatic, with the global average proportion of the elderly population reaching 9.8% in 2022, whereas the average for OECD countries stands at 18%. The proportion of the elderly population in Singapore is 19.1%, varying by area. Using the three values as thresholds, the 31 planning areas with available information on the elderly population were classified into 4 categories—superaged, aged, severely ageing, and mildly ageing—as in the base map in Figure A1 (Supplementary Section 2). Notably, eight superaged planning areas, with the share of the proportion of older people exceeding 20%, are situated primarily in central or southern Singapore. In contrast, many planning areas located at the urban fringe exhibit percentages ranging between 7% and 14%. There are no planning areas where the proportion of older people falls below 7%. Therefore, on the basis of the distribution of the planning area ranging from 7% to 14%, we classified them into two categories in the diagram: mildly ageing (7% < older proportion < 11.13%) and severely ageing (11.13% < older proportion < 14%).

The height indicates the accessibility of each type of AIP facility in the planning area: the higher the height, the more accessible the specific AIP facility is within that planning area. Specifically, we employed a gravity model to estimate accessibility, measured by the number of AIP facilities (as listed in Table A2) accessible within a 10-minute travel time (for detailed data, see Supplementary Section 3). Ideally, constructing AIP facilities should maximize efficiency by intentionally aligning with the spatial distribution of the elderly population. However, the results from the network analysis revealed practical spatial mismatches. Taking elderly-friendly services as an example, older adults in the Downtown Core—classified as 'severely aged'—have access to more than 10 such facilities within a 10-minute travel time. In contrast, those in the even more severely aged Ang Mo Kio region have access to only 0.54 facilities. According to an officer from the Singapore Land Authority (SLA), whom we interviewed, this mismatch arises primarily because SLA's current site-selection process for AIP facilities heavily relies on inter-departmental negotiations, lacking a quantifiable framework for determining optimal site selection. The framework proposed in this article aims to serve as a reference for addressing site-selection issues in future planning.

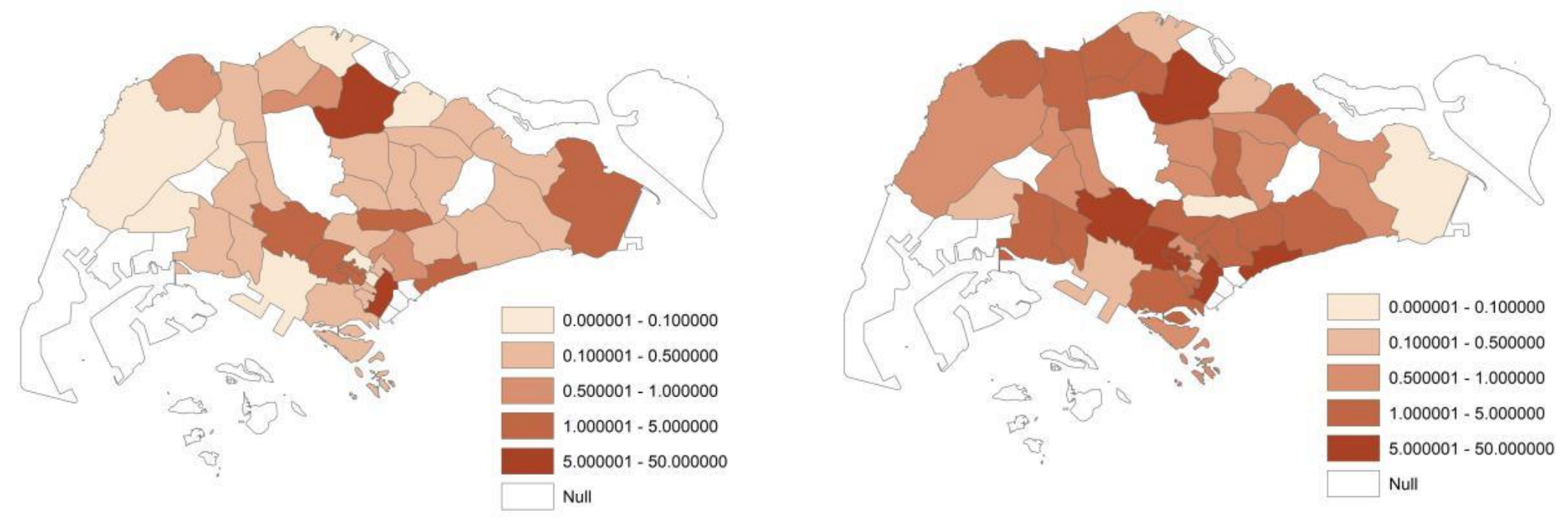

Parks　　　　Elderly-friendly services

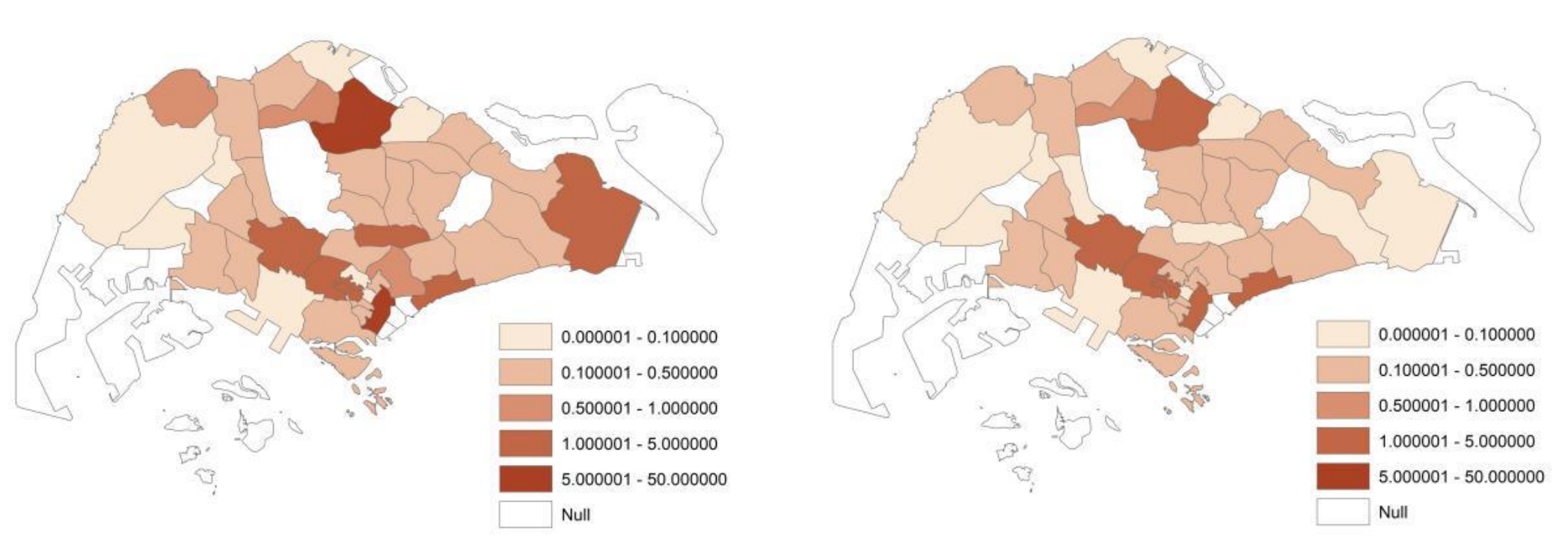

Community clubs　　　　Medical facilities

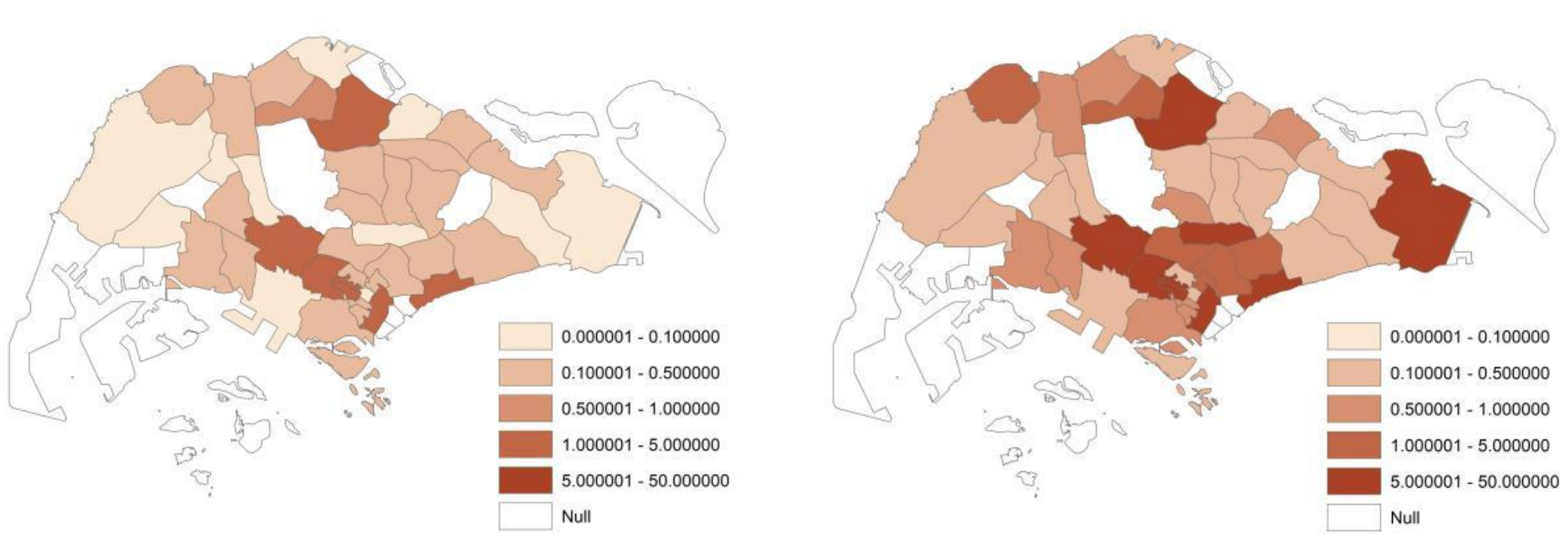

EDCC Hospitals

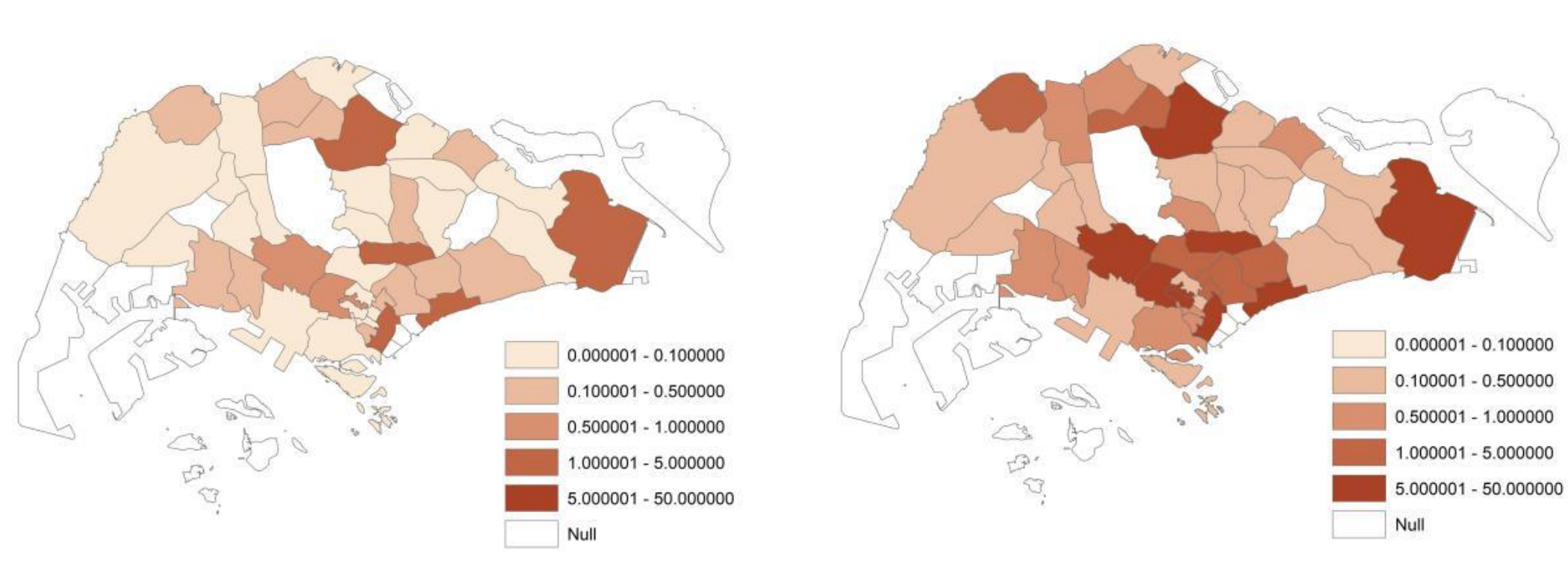

Healthier caterers Food courts

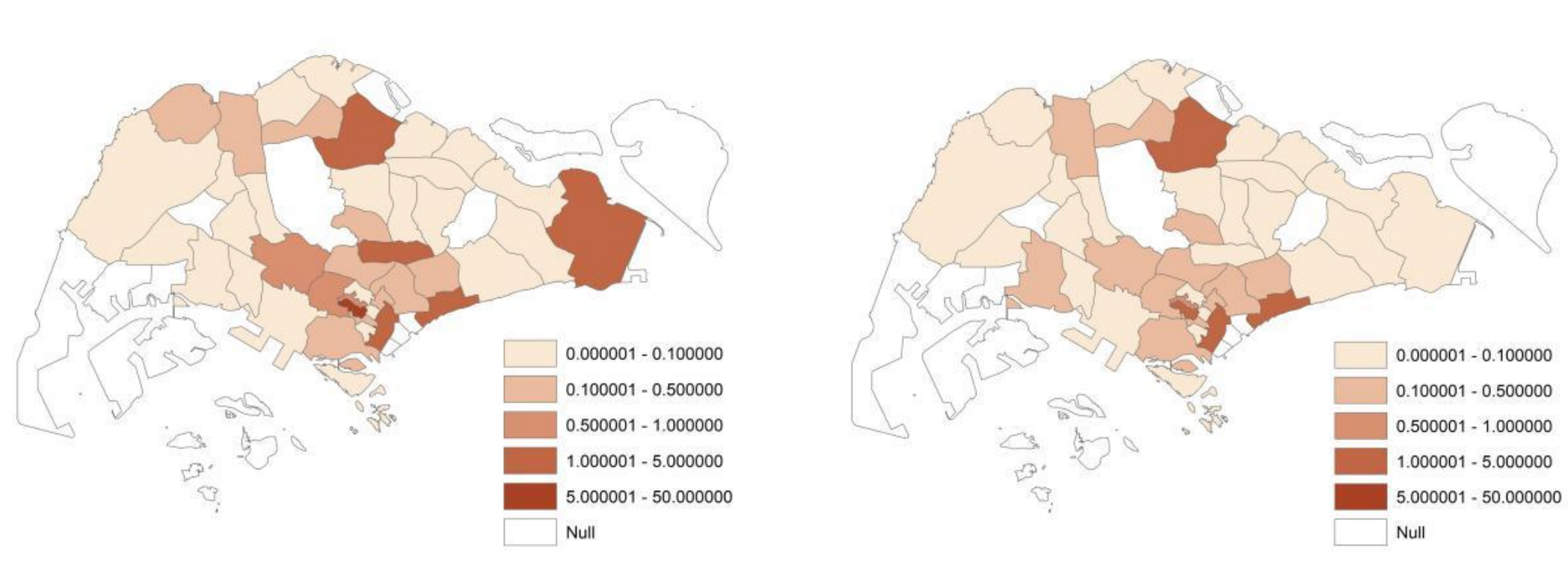

Supermarkets Disability services

Figure 3. Accessibility of facilities within a time distance of 10 minutes

Notes: *'Null' indicates no public housing in these planning areas. Darker red of the regional blocks indicate that more facilities can be reached within a 10-minute time distance.*

Singapore has made significant efforts to promote AIP practices. For example, the construction of new clinics outlined in the '*Action Plan for Successful Ageing*' was implemented from 2015 to 2020. When the negotiation costs among various departments are not considered, we find that newly constructed clinics can achieve Pareto improvement over five years with the aid of the quantitative assessment framework proposed in this paper. Figure 4 illustrates the geographic distribution of newly constructed clinics between 2015 and 2020. The areas with the lowest health facility accessibility did not have any planned construction from 2015 to 2020. With the aid of such method, urban planners can build new AIP facilities in the areas where they are most needed. This helps ensure more older people can benefit from the facilities.

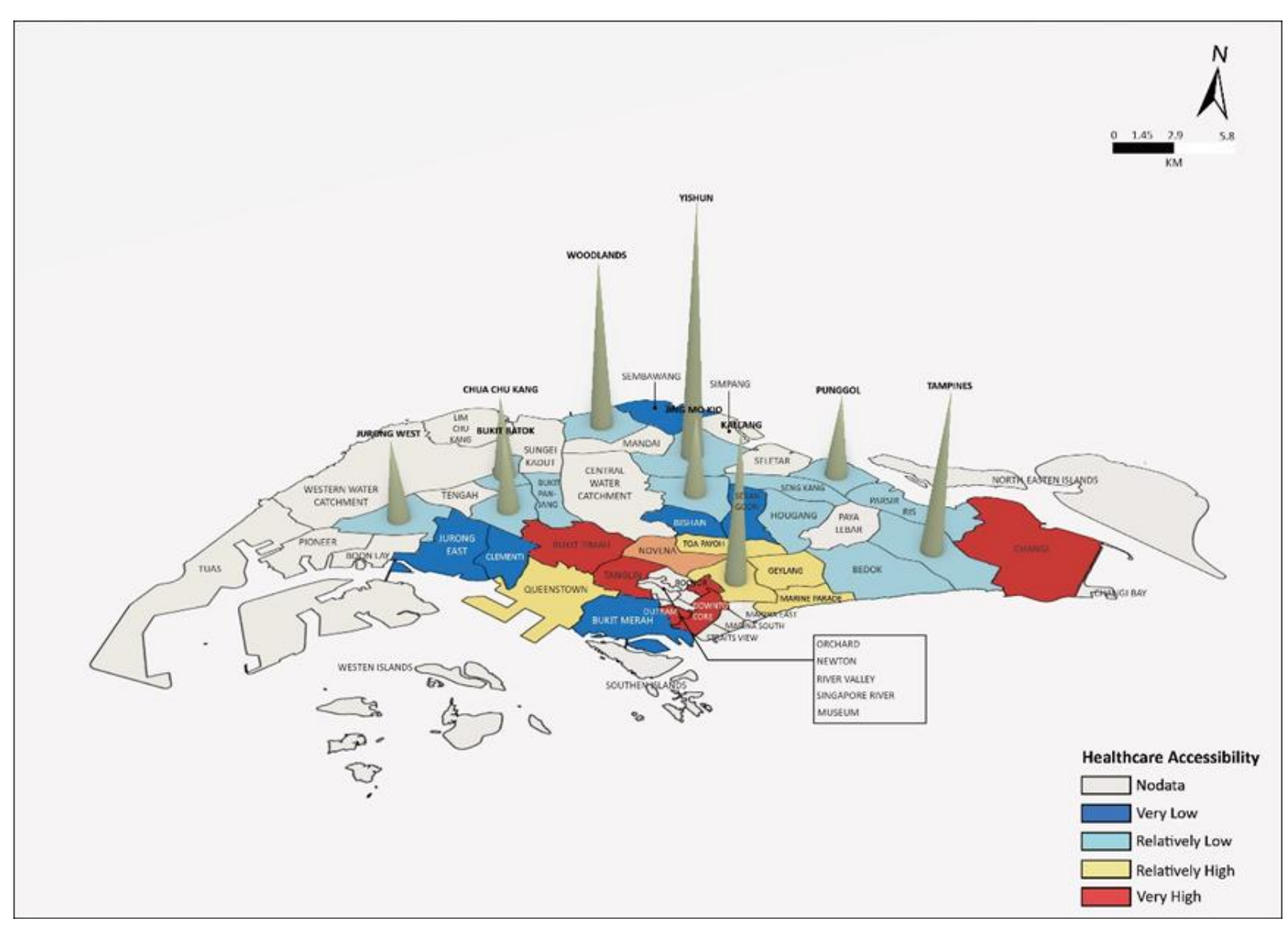

Figure 4. Number of new clinics between 2015 and 2020

Notes: *'No data' indicates no public housing in these planning areas. The four categories 'very low,' 'relatively low,' 'relatively high,' and 'very high' in Figure 4 represent the accessibility of medical facilities on the basis of percentile rankings. The planning areas represented by green cone-shaped structures indicate the presence of newly constructed medical facilities. The height of the cone represents the quantity of newly constructed medical facilities, with taller cones indicating more facilities.*

## *Qualifying the spatial matching degree: AFMI*

To further assess the spatial matching degree of AIP facilities in Singapore, in the second step, we developed an AIP facility matching index (AFMI) for each type of AIP facility by planning area, drawing upon the work of Holzer (1991) and Gobillon et al. (2007). The construction of the AFMI was based on the widely used dissimilarity index. The AFMI contributes novel insights by quantifying spatial mismatches and identifying underserved areas that conventional planning often overlooks. It reinforces the notion that regions with large ageing populations and high service demand frequently experience fragmented governance and a lack of integrated planning. By making these mismatches measurable, the AFMI offers a diagnostic tool that has been largely absent from ageing-related urban planning. Our findings address a longstanding gap between service needs and facility provision, promoting a more evidence-based and coordinated planning approach.

We first present a comprehensive assessment of AIP facility provision in each planning area, integrating the mismatches across all ten facility types within that area. The distribution of the integrated AFMI values across Singapore is shown in Figure 5, providing a holistic picture of how well each planning area is served overall. The darker colors indicate higher integrated AFMI values, reflecting a greater degree of mismatch. At the national level, we find an AFMI value of 0.24. Compared with the maximum value of 1, this value indicates a considerable degree of overall spatial mismatch in Singapore's AIP practices.

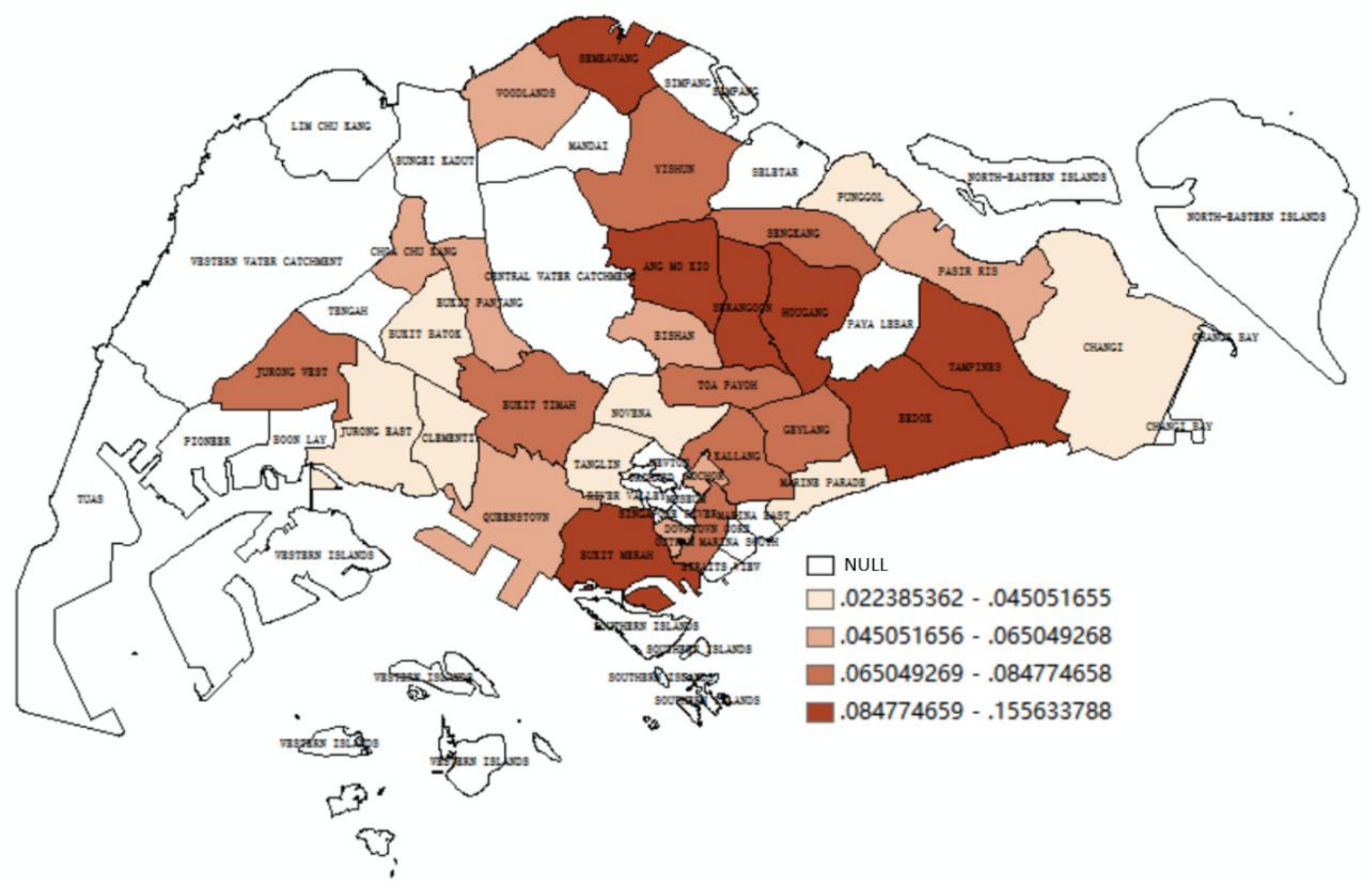

Figure 5. Spatial matching between the distributions of seniors and facilities: AIP facility matching indices

*Notes: 'No data' indicates no public housing in these planning areas.*

Subsequently, we present the AFMI values by grouping the ten AIP facility types into three broad categories—social venues, healthcare facilities, and general amenities—across planning areas with varying levels of population ageing. This categorization provides a new lens through which to understand facility demand, recognizing that active and dependent older adults may have differing needs across the three categories. The results are illustrated in Figure 6. Lastly, the breakdown of AFMI values for each individual facility type are presented in Table A3 in the Supplementary Information. Notably, there is considerable variation across facility types—for instance, food courts have a relatively high AFMI of 0.43, while medical facilities and community clubs report lower values of 0.13. Each of the three dimensions carries important policy implications, from offering a composite view of how well a planning area is served overall to pinpointing the single worst-performing facility type.

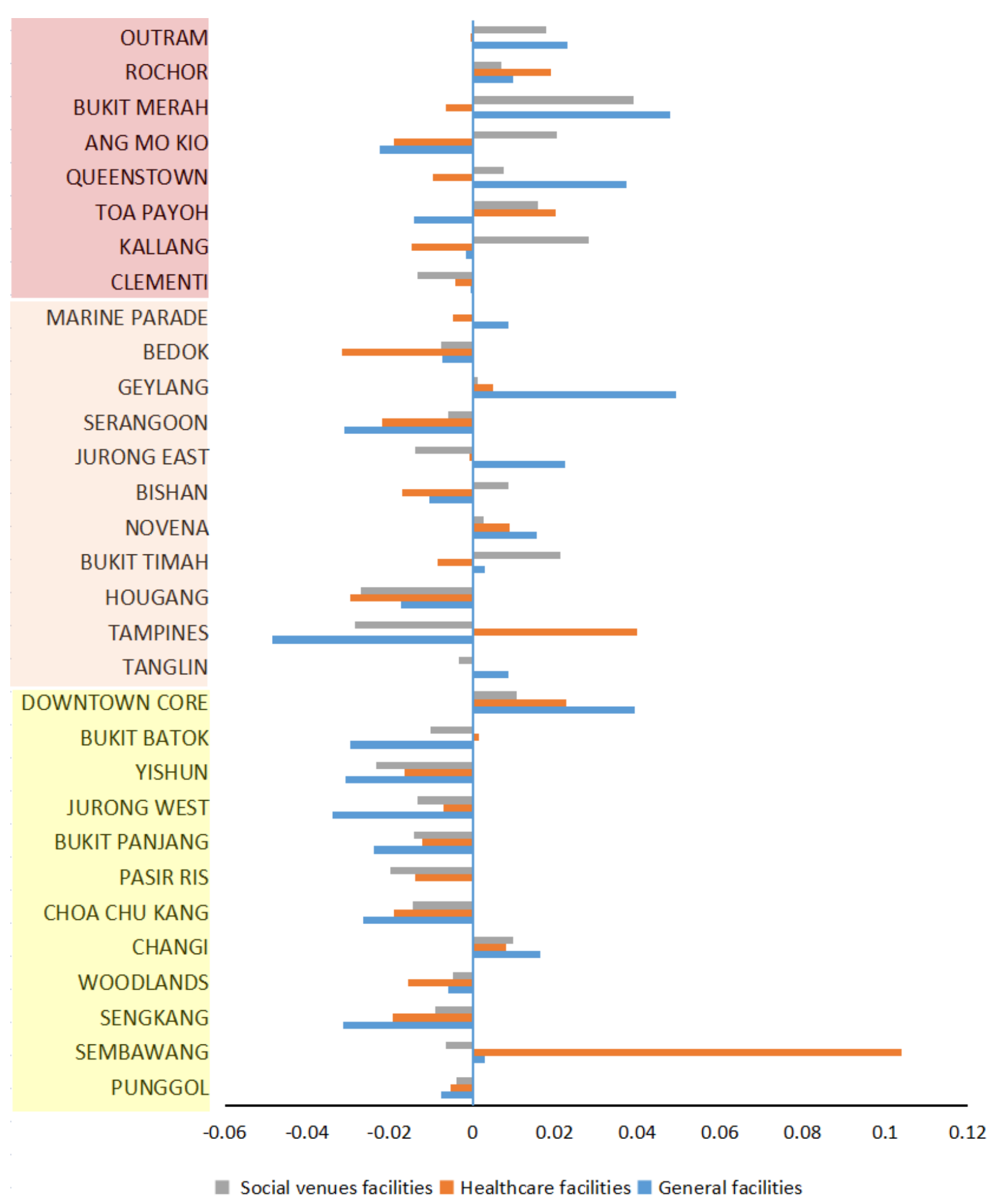

Figure 6. Spatial matching by category across the 31 planning areas

*Notes: Red, pink, and yellow indicate the superaged, aged, and ageing planning areas, respectively. Positive values (on the right) indicate oversupply, whereas negative values (on the left) indicate shortages. The length of the bar indicates the degree of severity.*

At the current level of supply, some facilities exceed the average level, whereas others fall short. We present the actual values (instead of absolute values) of the AFMIs for social venues, health care facilities, and general facilities separately across the 31 planning areas in Figure 5. An actual negative value indicates a shortage, and a positive value indicates oversupply. For example, an AFMI value of –0.02 indicates a moderate undersupply of a specific type of facility in a given planning area, relative to the national distribution of the elderly population. As a relative index, the sign denotes the direction of mismatch—negative values indicate relative undersupply, while positive values indicate oversupply. The magnitude reflects the severity of the mismatch: values closer to zero suggest a better alignment between facility provision and the spatial distribution of the elderly population. In this case, –0.02 implies that the facility allocation in the area is near optimal, though a slight shortfall remains. Details and an explanation of the AFMI equation can be found in the Methods section. The planning areas are arranged in descending order of the proportion of the population of older people, and they are categorized into superaged (marked in red on the y-axis), aged (marked in pink on the y-axis), and ageing (marked in yellow on the y-axis).

Interestingly, the evaluation results show that the superaged regions in Singapore, which are the policy target areas for the provision of AIP facilities, experience more oversupply than undersupply. On the other hand, the mildly aged planning regions tend to experience more undersupply than oversupply. The findings are of policy significance as such data basis can help urban planners to become more informed decisionmakers on the implementation of AIP facilities. Specifically, the AIP Facility Matching Index tells if there are any oversupplies or undersupplies of ageing-friendly facilities in a neighborhood or region, while the spatial network analysis can visualize the accessibility of AIP facilities across regions, which can aid the resource (re)allocation to (re)build AIP facilities at city level. It is important to notice that the AFMI is best applied at the city or region level. Extending the analysis to broader geographic scales may introduce noise in data, particularly in large countries with pronounced cross-regional heterogeneity in demographic and infrastructural conditions. In the U.S., for example, counties and census tracts are commonly used as standardized spatial units (Chyn & Daruich, 2025; Lou et al., 2023). In such contexts, the denominator should be defined in a context-sensitive manner: rather than relying on national totals, normalization within a state, metropolitan area, or other functionally relevant region may be more appropriate. Thus, to extend the AFMI to other contexts, it will be necessary to adjust it by ensuring that the external environmental characteristics of the broader area do not distort the density measures of older adults and elderly facilities.

In summary, the urban spatial analysis framework sets up a data-driven platform for policymakers and urban planners to make decisions in public resource allocation in building the AIP facilities or adjust decision directions timely in re-allocating resources across different regions.

Our findings imply that the AIP framework should take the spatial/environmental justice theory into consideration. In theory, it argues that it is important to reduce spatial inequality through policy interventions and urban planning, and to provide equal public services to ensure that all groups have equal access to spatial resources and opportunities (Moukouei et al., 2024; Venter et al., 2023; Jian et al., 2021), highlighting the need for a quantitative evaluation framework to assess the efficiency and effectiveness of an AIP practice.

## Discussion: Extension to Other Urban Settings

Singapore launched the Action Plan for Successful Ageing in 2015, with the core objective of enabling all older adults to age in place—a goal explicitly articulated in official policy documents. Led by a multi-agency taskforce, the plan encompassed multiple domains, including healthcare, community support, lifelong learning, and social participation. In total, over 70 initiatives were implemented. The government allocated approximately SGD 3 billion (USD 2.1 billion) to support the agenda and initiated the large-scale development of age-friendly infrastructure nationwide, such as Active Ageing Centres, senior-friendly transport systems, and barrier-free home modifications. Our study is situated within this institutional background and draws on spatial justice theories which emphasize justice in accessibility to urban opportunities (Moukouei et al., 2024). Specifically, we evaluate spatial justice in accessing AIP facilities, in light of the government's active promotion of age-friendly infrastructure.

In the same year, Amsterdam introduced a comparable ageing policy—the Amsterdam Ageing Agenda—and simultaneously joined the WHO Global Network for Age-friendly Cities and Communities. In contrast to Singapore's approach, Amsterdam focused on retrofitting existing community environments to improve the daily experiences of older residents. Key measures included installing rest benches with armrests, upgrading street lighting to warmer tones, and enhancing pedestrian accessibility. This strategy achieved notable success in promoting urban inclusiveness. However, the core of Amsterdam's model lies in small-scale, embedded modifications to existing public spaces rather than the construction of large-scale AIP facilities. Moreover, the city received partial funding through WHO support, and the overall financial investment remained relatively modest, relying more on community collaboration and incremental improvements led by local authorities.

Nonetheless, while age-friendly modifications enhance the convenience and safety of living environments for older adults, they do not constitute a comprehensive AIP system. These interventions primarily address spatial accessibility—the ability to live—but fall short of meeting more holistic needs related to care, healthcare, and social engagement—the ability to thrive. By contrast, Singapore's AIP model not only improves the physical environment but also integrates service delivery through Active Ageing Centres and community care networks, providing institutionalized and routine support for older residents. It is important to note, however, that this model requires substantial public investment. With the rollout of the Action Plan for Successful Ageing 2.0 (2023), Singapore gets to further optimize resource efficiency—for example, by locating AIP facilities at strategically selected, service-accessible nodes—to improve spatial/environmental justice (Venter et al., 2023; Moukouei et al., 2024).

To support forward-looking policy planning, we propose a two-pronged approach. First, urban planning should adopt data-informed indicators, such as facility-to-population ratios and spatial mismatch indicators (e.g., the Ageing Facility Mismatch Index (AFMI), Gravity model analyses) into regulatory frameworks and redevelopment priorities, including initiatives like *Action for All Ages 2.0*. Second, to overcome fragmented site selection practices, a cross-agency, quantitatively grounded decision-making framework should replace ad hoc negotiations. This includes regularly updating an AFMI-based site selection model to guide the equitable and efficient allocation of ageing-related infrastructure. It is important to note, however, that the current analytical framework does not incorporate land use regulations specific to each planning area. As such, final decisions should integrate AFMI results with local planning constraints and other contextual factors.

Our contributions are two-folded. First, we contribute to the evaluation of the AIP practice, by quantifying the (mis)match between population of the older people and the AIP facilities and place attributes. Moreover, our findings contribute to the reconceptualization of AIP theory by leading the way in incorporating environmental and spatial justice theory into the concept of building an AIP-friendly urban environment.

If a government aims to cover both active ageing population and dependent older people as much as possible, regardless of the city topology (e.g., compact or sprawling), the Singapore's AIP practice provides many learning points. However, the Singapore's AIP model is better suited for those land-constrained and densely populated cities, such as Hong Kong or Seoul, as the model is more spatially efficient in compact rather than expansive cities. Beyond ageing patterns and geographic characteristics, our cost-benefit analysis suggests that governments' fiscal capacity should also be incorporated into the assessment. In this context, a hybrid approach that integrates elements of the Singapore model with other frameworks is recommended, tailored to the varying fiscal, environmental, and demographic characteristics of different cities.

For example, in a low-density city with less centralized government intervention in ageing care, such as Vancouver, a 'Singapore–San Francisco' hybrid model may be more appropriate. The San Francisco model focuses on providing self-funded elderly care services at centralized locations for higher-income elderly groups, while the Singapore model balances the elderly care needs of different income groups across neighborhoods, albeit with significant government investment. This 'Singapore–San Francisco' hybrid leverages the strengths of both approaches and can be generalized to fiscally well-resourced, less land-constrained cities, such as Amsterdam, where a balanced model that combines government coordination with market-driven elderly care proves advantageous.

The Tokyo model, with its emphasis on government-funded home care and comprehensive healthcare support, combined with Singapore's spatial efficiency and comprehensive neighborhood coverage, can enhance the welfare of all ageing groups. Such a hybrid model is particularly well-suited for densely populated, fiscally robust cities facing significant ageing challenges—such as New York, Toronto, and Hong Kong—where the integration of government-funded healthcare and efficient spatial planning is essential due to limited land and high service demands.

## Conclusion

In conclusion, this study demonstrates a new AIP model using Singapore as an example, drawing insights from existing AIP models. Singapore's AIP model supports older people, especially independent people, to actively age in their own homes in elderly-friendly communities with accessible facilities and neighbourhood services. It offers accessibility and cost-effectiveness for older people. We also propose an evaluation framework that incorporates spatial network analysis and the AFMI to quantify the degree of spatial matching, which can be applied to other cities.